\documentclass[letterpaper]{jpconf}

\newcommand{\gcc}{\mathrm{g~cm^{-3} }}
\newcommand{\cms}{\mathrm{cm~s^{-1}}}

\renewcommand\abovecaptionskip{5pt}

\usepackage{graphicx}

\begin{document}
\title{The Physics of Flames in Type Ia Supernovae}

\author{M.~Zingale$^1$, S.~E.~Woosley$^1$, J.~B.~Bell$^2$, M.~S.~Day$^2$, C.~A.~Rendleman$^2$}

\address{$^1$ Department of Astronomy and Astrophysics, University of California, Santa Cruz, Santa Cruz, CA 95064}
\address{$^2$ Center for Computational Science and Engineering, Lawrence Berkeley National Laboratory, Berkeley, CA 94720}

\ead{zingale@ucolick.org}

\begin{abstract}

We extend a low Mach number hydrodynamics method developed for
terrestrial combustion, to the study of thermonuclear flames in Type
Ia supernovae.  We discuss the differences between 2-D and 3-D
Rayleigh-Taylor unstable flame simulations, and give detailed
diagnostics on the turbulence, showing that the kinetic energy power
spectrum obeys Bolgiano-Obukhov statistics in 2-D, but Kolmogorov
statistics in 3-D.  Preliminary results from 3-D reacting bubble
calculations are shown, and their implications for ignition are
discussed.  
\end{abstract}

\section{Introduction}

Type Ia supernova (SNe Ia) are believed to result from the
thermonuclear explosion of a Chandrasekhar mass white dwarf.  A
thermonuclear flame, born at or near the center of the white dwarf
must accelerate tremendously, approaching the speed of sound, in order
to account for observational constraints.
Uncertainties remain at all stages of the process, from how it
ignites, to how the flame accelerates, to whether it may at
some point transition to a detonation (see
\cite{hillebrandtniemeyer2000} for a review).  Speculation on the
latter has suggested \cite{khokhlov:1991,niemeyerwoosley1997} that the
optimal conditions for a transition to detonation to occur are when
the flame enters the distributed burning regime, at densities of $\sim
10^7~\gcc$, although other mechanisms have recently been proposed
\cite{gcd}.

The range of relevant length scales in the white dwarf is
enormous---from the $10^8$~cm radius of the white dwarf down to the
$10^{-4}$--$10^1$~cm thickness of the flame.  No single simulation can
encompass all length scales, so approximations must be made.  Large
scale 3-D simulations (e.g.~\cite{roepke2005,gamezo:2003}) put the
entire star on the grid and resolve scales down to $\sim 10^5$~cm,
and use a subgrid model to describe the physics of the turbulent
burning on the small scales.  We have chosen to take a complimentary
approach, investigating the physics of turbulent thermonuclear burning
on the flame scales, with the hope to use what we've learned to push
up to large scales.

Over the last few years, as part of a collaboration between the SciDAC
Supernova Science Center at UCSC and the Center for Computational
Science and Engineering at LBL, we've introduced a low Mach number
model for simulating thermonuclear flames in SNe Ia \cite{Bell:2004}.
In the low Mach number limit, the pressure is decomposed into a
thermodynamic component, $p_0$, and an ${\cal O}(M^2)$ dynamic
component, $\pi$.  This leads
to the set of low Mach number thermonuclear flame equations
\begin{eqnarray*}
\frac{\partial (\rho U)}{\partial t} + \nabla\cdot(\rho UU)
 &=& -\nabla \pi + \rho \vec g \enskip , \\
\frac{\partial(\rho h)}{\partial t} + \nabla\cdot(\rho U h )
&=& \nabla\cdot(\lambda \nabla T) - \sum_k\rho q_k\dot\omega_k \enskip ,\\
\frac{\partial (\rho X_k)}{\partial t} + \nabla\cdot(\rho  U X_k)
&=& \rho\dot\omega_k \enskip .
\end{eqnarray*}
where $\rho$, $U$, and $h$ are the density, velocity, and enthalpy
respectively.  The enthalpy is related to the internal energy, $e$
through $h = e + p/\rho$.  $X_k$ is the abundance of the
$k^\mathrm{th}$~isotope, with reaction rate $\dot\omega_k$ and energy
release $q_k$.  $T$ is the temperature and $\lambda$ is the thermal
conductivity.  Finally, $\vec g$ is the gravitational acceleration.
This system is constrained to evolve at constant pressure, $p_0$,
and differentiating the pressure along particle paths yields
an elliptic equation for the velocity:
\begin{equation}
\nabla\cdot U = \frac{1}{\rho\frac{\partial p}{\partial \rho}}
\left(
\frac{1}{\rho c_p}
\frac{\partial p}{\partial T}\left(
  \nabla\cdot\lambda \nabla T - \sum_k\rho \left(q_k + 
\frac {\partial h}{\partial X_k}
\right)\dot\omega_k\right)
+ \sum_k\frac{\partial p}{\partial X_k}\dot\omega_k
\right)  .
\label{eq:sn_constraint}
\end{equation}
This system filters out sound waves, allowing for much larger
timesteps than a compressible code.  Eq.~(\ref{eq:sn_constraint}) is
similar to the incompressible constraint, with sources
representing thermal diffusion and energy generation across the
flame. The solution method is based on second-order projection methods
for incompressible flow, embedded in
an adaptive mesh refinement framework.  Further details of our
derivation of the low Mach number model for nuclear flames are given
in~\cite{Bell:2004}.

\section{Results}
 
We have applied this numerical model to several multidimensional
studies of thermonuclear flame propagation in SNe Ia
\cite{SNld,SNrt,SNrt3d}.  Our approach to date has been to resolve the
thermal structure of the flame numerically, eliminating the need for a
flame model.  The propagation of the flame is entirely determined by
the input physics (reaction rate, conductivity) and the flow field.
This constrains the regimes we can study however, as we need to have
both the flame thickness resolved and several unstable wavelengths
contained on our grid.  

\begin{figure}[t]
\includegraphics[width=3.82in]{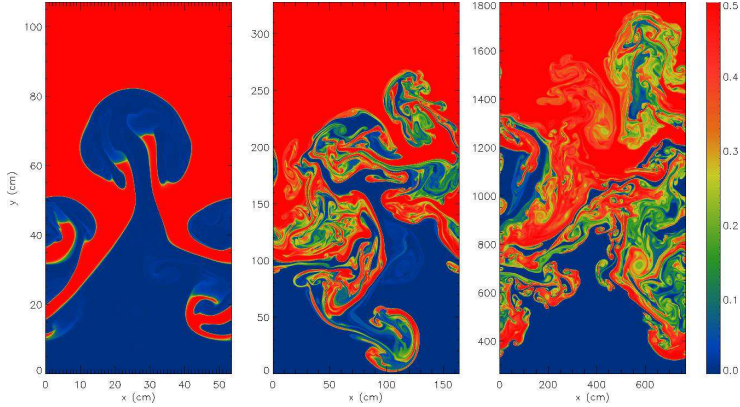}\hspace{0.1in}
\begin{minipage}[b]{2.2in}
\caption{\label{fig:2d} 2-D RT unstable carbon flames at
$1.5\times 10^7~\gcc$ (left), $10^7~\gcc$ (middle), and $6.67\times
10^6~\gcc$ (right), showing the carbon mass fraction.  As the density
decreases, RT dominates over the burning and we enter
the distributed burning regime \cite{SNrt}. \\}
\end{minipage}
\end{figure}

\subsection{Rayleigh-Taylor generated turbulence}

Fig.~\ref{fig:2d} shows 2-D Rayleigh-Taylor (RT) unstable flames at
three different densities~\cite{SNrt}.  At the highest density, the
reactions proceed quickly, suppressing the RT instability on the small
scales.  At the low density end, the small modes of the RT instability
grow quickly, and the reactions are unable to burn them away.  Here,
modes smaller than even the flame thickness itself are unstable, and a
large mixed region of fuel and ash appears.  This is the beginning of
the distributed burning regime.  In all cases, the RT instability
greatly wrinkles the flame, creating more surface area, and
dramatically accelerating the flame.

\begin{figure}[t]
\includegraphics[width=2.2in]{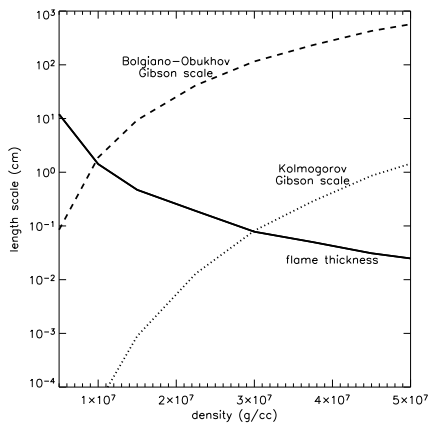}
\hspace{0.24in}
\begin{minipage}[b]{3.8in}
\caption{\label{fig:regimes} Turbulent flame properties in the
supernova.  Shown are the flame thickness (solid), the Gibson scale
assuming Kolmogorov scaling (dotted line), and the Gibson scale
assuming BO statistics (dashed line).  The transition to
distributed burning, when the Gibson scale is less than the flame
thickness, is smaller for BO scaling than for Kolmogorov
scaling, as first discussed in \cite{niemeyerkerstein1997}.  The flame
parameters for this figure were computed with the low Mach number 
algorithm discussed above and a single, screened $^{12}$C + $^{12}$C
reaction rate.\\ }
\end{minipage}
\end{figure}

The nature of the turbulence cascading down to the flame scale will
directly affect the density at which we enter the distributed burning
regime.  It was argued that the RT instability would lead to a
potential energy cascade, and that Bolgiano-Obukhov (BO) statistics
($u(l) = u^{\prime} (l/L)^{3/5}$) should be used
\cite{niemeyerkerstein1997}, leading to a lower transition density
than Kolmogorov statistics ($u(l) = u^{\prime} (l/L)^{1/3}$).
Fig.~\ref{fig:regimes} illustrates this by computing the Gibson scale
for both statistics as a function of density---this is the scale at
which the flame can burn away a turbulent eddy before it can turnover.
Here we have assumed that the integral scale, $L$, is $10^6$~cm and
the turbulent velocity on that scale, $u^{\prime}$, is $10^7~\cms$,
consistent with the numbers presented in \cite{niemeyerwoosley1997}.
If the transition is at a lower density, as dictated by BO scaling, it
was suggested that a deflagration-detonation transition (DDT)
initiated in the distributed burning regime would be more difficult
\cite{niemeyerkerstein1997}, due to the density sensitivity of the
carbon reaction rates.  Our 2-D RT studies \cite{SNrt} also suggest
that a DDT is not possible.  The true nature of the turbulent cascade
can be determined through resolved flame simulations.

\begin{figure}[t]
\centering
\includegraphics[height=2.1in]{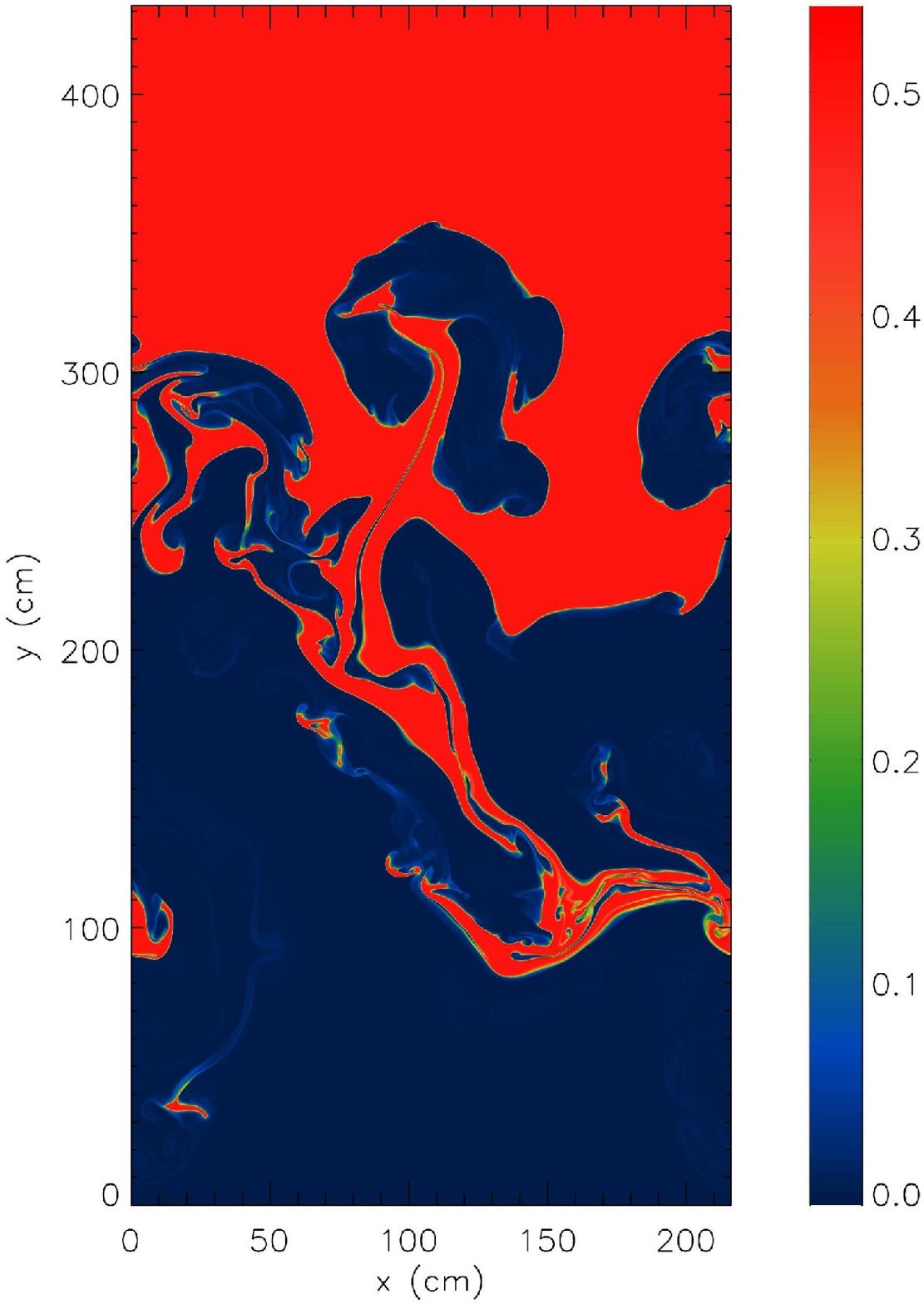}
\hskip 3 mm
\includegraphics[height=2.1in]{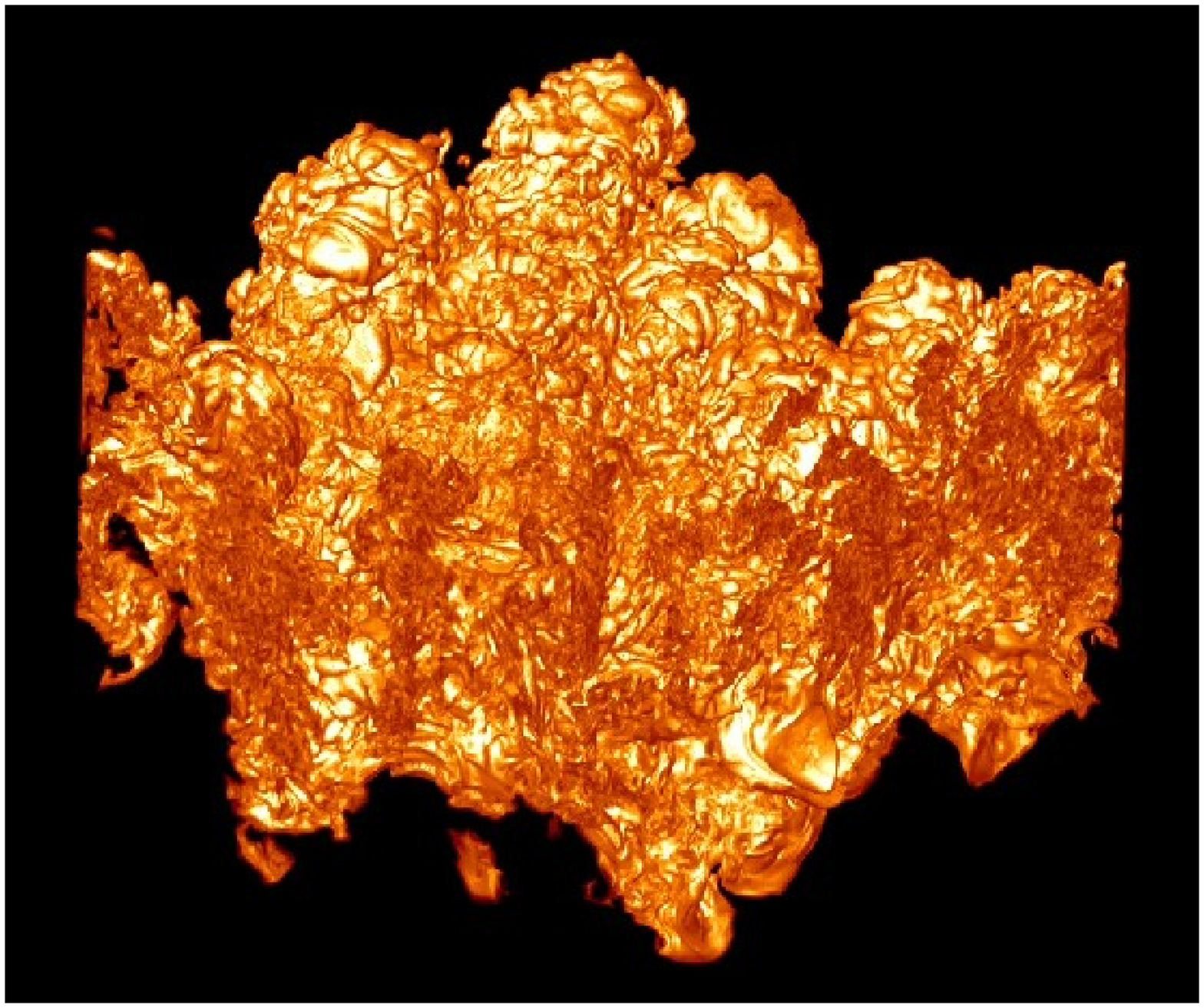}
\hskip 3 mm
\includegraphics[height=2.1in]{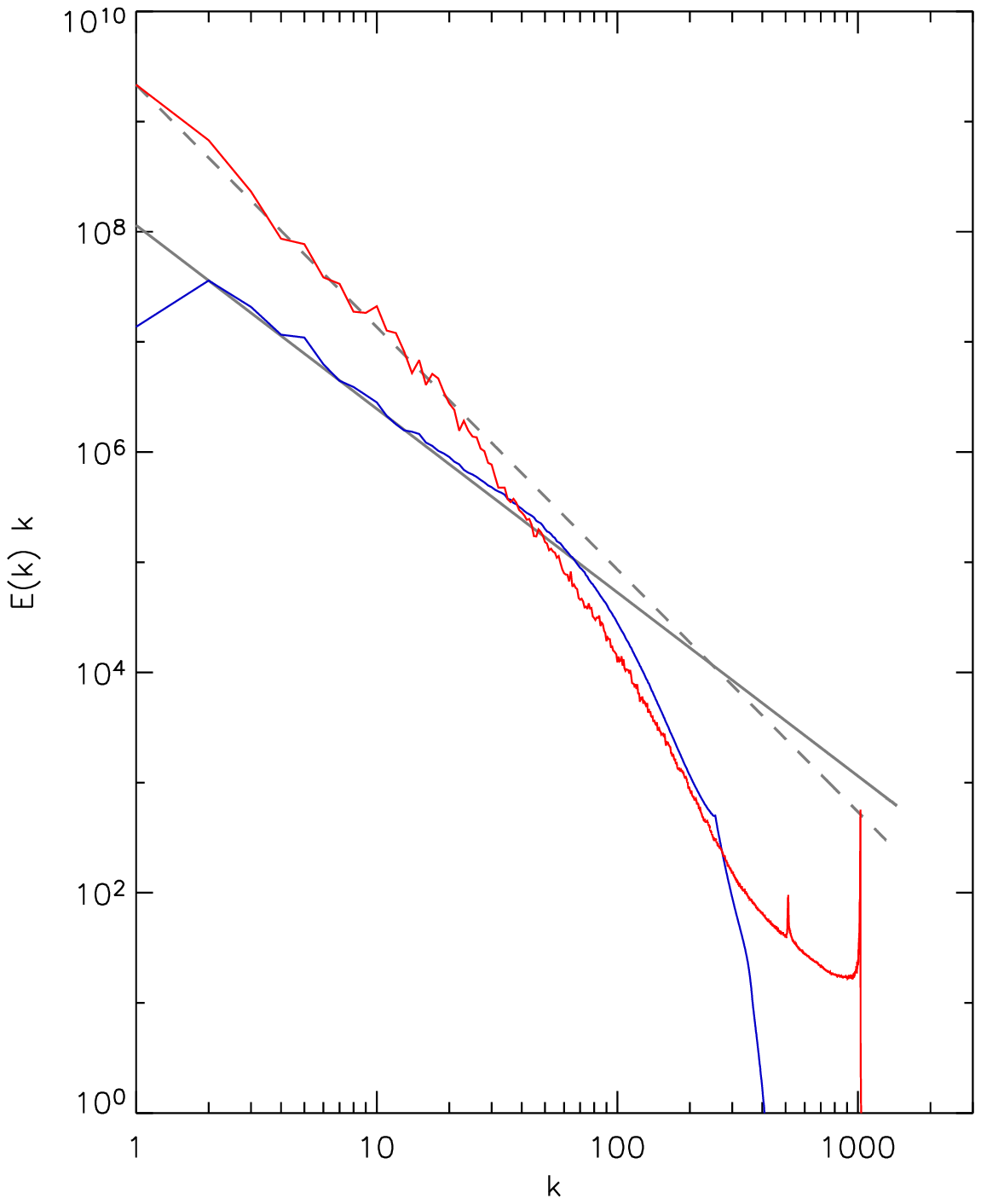}
\caption{ \label{fig:3d} 2-D (left) and 3-D \cite{SNrt3d} (center) RT unstable
carbon flame at $1.5\times 10^7~\gcc$.  The kinetic energy power
spectrum (right) shows the 3-D flame (blue line) following a $-5/3$ power
law (gray solid) and the 2-D flame (red line) following a $-11/5$ power
law (gray dashed).  The 2-D simulation was run in a large domain
(216~cm wide vs.\ 53.5~cm wide for the 3-D), necessary to see the
scaling trend.  This wider domain accounts for the higher peak kinetic
energy.}

\end{figure}

We studied a 3-D RT unstable flame in detail \cite{SNrt3d}, and found
that after the linear growth, turbulence dominates the dynamics.
Fig.~\ref{fig:3d} shows a volume rendering of the carbon mass fraction
and the associated kinetic energy power spectrum.  We see that it is a
Kolmogorov power spectrum, not BO, as previously suggested
\cite{niemeyerkerstein1997}.  Gravity creates a preferred direction in
the domain, and the turbulence is strongly anisotropic on the large
scales.  However, on the small scales, the turbulence becomes more and
more isotropic, as determined by looking at isosurfaces of the Fourier
transform of the turbulent kinetic energy in $\bf{k}$-space
\cite{SNrt3d}.

Recent analytic work \cite{chertkov2003} argued that BO statistics
should only apply in 2-D.  Fig.~\ref{fig:3d} also shows a 2-D reactive
RT instability, and the power spectrum shows a clear $-11/5$ scaling
over more than a decade in wavenumber.  It is essential to use a large
domain, with many unstable modes to see this scaling in 2-D---this 2-D
simulation is effectively $2048\times 4096$ zones.  This comparison
between 2-D and 3-D confirms the assertions in \cite{chertkov2003},
and therefore, for SNe Ia, 3-D turbulence models should assume Kolmogorov
scaling, not BO.  With this dimensionality dependence, together with
Fig.~\ref{fig:regimes}, we can understand why our 2-D study
\cite{SNrt} showed a lower transition density to distributed burning
than our corresponding 3-D RT flame \cite{SNrt3d}.  This difference in
RT generated turbulence demands that any SNe Ia simulations be run in
3-D, since turbulence acts on all scales of the problem.  Since the
transition density is higher with Kolmogorov turbulence, it may be
easier to ignite a detonation than our 2-D estimates \cite{SNrt} have
suggested.  Studies of flames interacting with Kolmogorov turbulence
in 3-D are underway, and will allow us to explore more fully whether a
transition to detonation is possible.


\subsection{Bubbles}

Finally, we consider an alternate configuration from the flame sheets
that we have been studying, a burning bubble.  The
ignition of an SNe Ia likely begins with one or many hotspots that
begin to burn faster than they can cool by expansion.  These bubbles
are buoyant and rise as they grow, possibly merging as they consume
the carbon fuel in the star.  The drag force on a rising bubble is
smaller than that on a sheet, so we would expect different dynamics to
ensue.  Of particular interest is the small scale structure that
develops on the bubble.  The exploding white dwarf has an extremely
large Reynolds number, so we expect turbulence to dominant early in
the evolution of the reacting bubble.  It is possible that this
turbulence will shed sparks from the bubble as it rises, and if these
are small enough, they can be entrained in the ambient flow instead of
buoyantly rising on their own, igniting the star in new regions.

\begin{figure}[t]
\includegraphics[width=3.in]{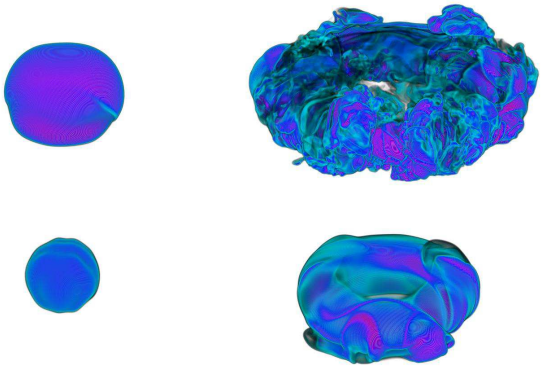}\hspace{0.28in}
\begin{minipage}[b]{3.0in}
\caption{\label{bubble} 3-D simulation of a buoyant rising bubble at
$1.5\times 10^7~\gcc$, at four different times.  The reactions are
slow enough at this density that the initially spherical bubble
deforms into a torus due to shear at the top.  At late times, the
bubble is becoming turbulent, with lots of small scale structure
appearing.  \\}
\end{minipage}
\end{figure}

Fig.~\ref{bubble} shows the preliminary results of a burning rising
bubble.  This was again done at a density of $1.5\times 10^7~\gcc$,
far from the conditions of ignition, but approachable by DNS
calculations.  The general features of the evolution should be
qualitatively the same as those of a bubble near ignition conditions.
This is an exceptionally large calculation, corresponding to a
$2048\times 2048\times 3072$ zone effective grid.  At present, over
200 million zones are carried, and this calculation is still in
progress.  As the figure shows, at late times, a hole develops in the
center of the bubble, due to the strong shear at the top of the
bubble, which suppresses the burning.  At higher densities, the faster
reaction rate may suppress this hole formation.  Also apparent is the
growth of small scale structure on the sides of the bubble, driven by
shear.  This is only seen in high resolution 3-D studies of reacting
bubbles.  Large scale simulations employing flame models would likely
wash this structure away, preventing the shedding of sparks, and
perhaps missing a key component of the ignition process.  More studies
of resolved burning rising bubbles at high Reynolds number are needed
to determine how large of an effect this spark shedding may be.

\section{Summary}

We presented results from a program of study of small-scale flame
physics.  We showed that in 3-D, the turbulent cascade obeys
Kolmogorov scaling while in 2-D, BO scaling is observed.  This
difference in turbulence scaling demands that SNe Ia simulations be
run in 3-D.  It is interesting to note that the Sharp-Wheeler model
for RT \cite{sharp1984} predicts a speed, $u \sim l^{1/2}$, between
BO and Kolmogorov scaling.  Flame models using different assumptions
of the turbulence speed on small scales will lead to differing
results.  If large scale simulations accurately capture the turbulent
cascade for more than a decade in wavenumber, then it is a good
approximation that the turbulence fed into the subgrid model is
Kolmogorov and isotropic.

Of critical importance to understanding the early stages of the
explosion is the dynamics of reacting bubbles.  Resolved calculations
are very difficult, but can have enormous consequences.  If the
bubbles quickly become turbulent, as expected in the high Reynolds
number flow in the white dwarf, then it is likely that sparks will be
shed and advect with the background convective motions to ignite other
regions of the star.  This may make multipoint, time-dependent
ignition an essential ingredient in the explosion process.

Current work involves extending our low Mach number model to
accommodate multiple scale heights.  This would allow for full star
calculations, while retaining the ability to take timesteps restricted
by only the fluid velocity rather than the sound speed.  This would be
ideal for studies of the ignition process itself, and allow for the
evolution to continue directly into the explosion phase.  These large
scale simulation will be the focus of the next years of our
collaboration.

\ack

Support for this work was provided by DOE grant No.\ DE-FC02-01ER41176
to the Supernova Science Center/UCSC and the Applied Mathematics
Program of the DOE Office of Mathematics, Information, and
Computational Sciences under contract No.\ DE-AC03-76SF00098.  Work
presented used the NERSC seaborg, ORNL cheetah,
and the NASA Ames Columbia machines.

\section*{References}

\end{document}